\documentclass[a4paper]{article}

\usepackage[algosection]{algorithm2e}
\usepackage{amsfonts,amsmath}
\usepackage{amssymb}
\usepackage{amsthm}
\usepackage[affil-it]{authblk}
\usepackage{booktabs}
\usepackage{comment}
\usepackage[update]{epstopdf}
\usepackage{fancyvrb}
\usepackage[final]{graphicx}
\usepackage{hyperref}
\usepackage{import}
\usepackage{latexsym}
\usepackage[final]{listings}
\usepackage{multirow}
\usepackage[numbers]{natbib}
\usepackage{paralist}
\usepackage{pgfplots}
\usepackage[indention=10pt,singlelinecheck=false,caption=false,nearskip=0pt,farskip=0pt,aboveskip=0pt,captionskip=2pt]{subfig}
\usepackage{tikz}
\usepackage[colorinlistoftodos,shadow,obeyDraft,textsize=small]{todonotes}
\usepackage{url}
\usepackage{xspace}

\usepgflibrary{plothandlers}
\usetikzlibrary{plotmarks}
\makeatletter
\renewcommand\bibsection%
{
\section*{\refname
\@mkboth{\MakeUppercase{\refname}}{\MakeUppercase{\refname}}}
}
\makeatother

\graphicspath{{images/}}

\newtheorem{definition}{Definition}

\newcounter{rulecounter}
{\itshape}{\rmfamily}

\newcounter{querycounter}
\newtheorem{queryEnv}{Query}{\itshape}{\rmfamily}

\newcommand{\tuple}[1]{\langle #1 \rangle }

\newcommand{\set}[1]{\ensuremath{\mathbf{#1}}}

\urldef{\mailsa}\path|{firstname.lastname}@insight-centre.org|
\urldef{\mailsb}\path|{firstname.lastname}@wu.ac.at|

\begin{document}

\title{Query Based Access Control for Linked Data}

\author[1]{Sabrina Kirrane}
\author[2]{Alessandra Mileo}
\author[1]{Axel Polleres}
\author[3]{Stefan Decker}

\affil[1]{Vienna University of Economics and Business, Austria}
\affil[2]{Dublin City University, Ireland}
\affil[3]{Fraunhofer FIT, RWTH Aachen University, Germany}


\date{}
\maketitle

\begin{abstract}
%As a result there is a need for models and frameworks that allow data publishers to control access to their data. 
%Although a number of access control models and frameworks have been put forward, the security implications arising from granting access to partial data remains an open issue. Furthermore, very little research has been conducted into the correctness of the potential solutions. 
%and examine the security implications associated with several SPARQL 1.1 query constructs. In addition, 
% and demonstrate how model checking can be used to ensure that the proposed query rewriting algorithm is secure, sound and maximum. 

In recent years we have seen significant advances in the technology used to both publish and consume Linked Data. However, in order to support the next generation of ebusiness applications on top of interlinked machine readable data suitable forms of access control need to be put in place.
Although a number of access control models and frameworks have been put forward, very little research has been conducted into the security implications associated with granting access to partial data or the correctness of the proposed access control mechanisms. Therefore the contributions of this paper are two fold: we propose a query rewriting algorithm which can be used to partially restrict access to SPARQL 1.1 queries and updates; and we demonstrate how a set of criteria, which was originally used to verify that an access control policy holds over different database states, can be adapted to verify the correctness of access control via query rewriting. \end{abstract}

% \begin{keywords}
% Access Control,SPARQL, RDF, Linked Data
% \end{keywords} 

\section{Introduction}
The term Linked Data Web (LDW) is used to describe a World Wide Web where data is directly linked with other relevant data using machine-accessible formats \cite{OHara2013, Heath2011}. 
%Many of the technologies required to realise the LDW vision are already in place, for example RDF, RDFS, OWL and SPARQL. 
Although the LDW has seen considerable growth in recent years, the focus continues to be on linking public data. This can partially be attributed to the fact that no formal recommendation exists for allowing partial access to Linked Data based on predefined access control policies.
%for the secure querying of the LDW.  

Several researchers have proposed access control strategies for the Resource Description Framework (RDF), which could be applied to Linked Data. 
Broadly speaking, these frameworks enforce access control either at the data layer \cite{Dietzold2006, muhleisen2010swrl}, the query layer \cite{Abel2007, Franzoni2007, Chen2009, oulmakhzoune2010,oulmakhzoune2012} or a combination of both \citep{li2008}.
%
%Broadly speaking access control frameworks for RDF data enforce access control either at the data layer \cite{Jain2006, Papakonstantinou2012}, the query layer \cite{Costabello2012, Sacco2011, Kirrane2013} or a combination of both \cite{Abel2007}.
% motivate our focus on query-based:
Enforcement of access control at the data layer is concerned with removing unauthorised data from a dataset. 
%Whereas, enforcement at the query layer relies on query rewriting techniques to ensure that the query results are restricted to those that are permitted by an access control policy.
Whereas, enforcement at the query layer relies on using query rewriting techniques to ensure that only authorised query results are returned.
%
%Regardless of the proposed enforcement mechanisms, to date researchers have focused on evaluating the performance of their access control systems as opposed to verifying the correctness of their access control algorithms. 
 
%Although the enforcement of access control at the query layer caters for the specification of policies at multiple levels of granularity 
%
%When it comes to allowing access to partial query results, it is possible to use a filtering algorithm to generate a dataset which does not contain any prohibited data, and to subsequently execute the query against the filtered dataset. Alternatively it is possible to develop a query rewriting algorithm that can be used to ensure that unauthorised data is not returned by the query.
%
%Given the potential scalability issues associated with results filtering,
%When access control is enforced using results filtering, it is necessary to create a temporary dataset for each user, which would have consequences from a scalability perspective, 

Given we need to cater for many users, with many different authorisations a filtering approach will not scale well, therefore in this paper we adopt a query rewriting approach. 
To date researchers have proposed query rewriting strategies that involve adding contextual information pertaining to the requester to the query using path expressions and bindings~\cite{Abel2007, Franzoni2007,oulmakhzoune2012}, or adding bindings for authorised/unauthorised classes, properties or instances using filters~\cite{Chen2009,oulmakhzoune2010}. 
%
%When it comes to access control for Linked Data, in the case of the former the dataset may not contain information that can be used to control access. 
%In this paper we demonstrate how SPARQL 1.1 queries and updates can be rewritten so that they behave as if the unauthorised data is not present in the dataset.
Our query rewriting strategy builds on the latter approach by demonstrating how quad patterns can be used to grant or deny access to RDF data at multiple levels of granularity (triple, named graph, classes, properties, instances).
In addition, we demonstrate how our query rewriting strategy can be used not only to enforce access control over basic graph patterns but also in conjunction with SPARQL 1.1 queries that include subqueries and negation, graph update operations and graph management operations.
%In addition, we demonstrate how both SPARQL 1.1 queries and updates can be rewritten so that the query behaves as if the unauthorised quad pattern is not present in the dataset.
%In addition, we demonstrate how both SPARQL 1.1 queries and updates can be rewritten so that the query behaves as if the unauthorised quad pattern is not present in the dataset.
%only authorised data is returned to the requester. 
Although in this paper we discuss how access can be restricted to a single quad pattern as we use \texttt{FILTER NOT EXISTS} to deny access to unauthorised data our approach can easily be extended to work with graph patterns.
In order to verify the effectiveness of our query rewriting strategy, we define a set of criteria that can be used to compare the results obtained when a query is executed over a dataset where all unauthorised data has been removed, and the results obtained when a query is rewritten and executed over a dataset, which contains both authorised and unauthorised data. 
The contributions of the paper can be summarised as follows: we
\begin{inparaenum}[(i)]
\item formally define a set of criteria which can be used to verify the correctness of access control via query rewriting;
\item propose query rewriting strategies that can be used to partially restrict access to SPARQL 1.1 queries and updates; and
\item demonstrate how the proposed correctness criteria can be used to verify existing query rewriting strategies.
%\item verify the correctness of our access control algorithms using the proposed correctness criteria.
\end{inparaenum}

The remainder of the paper is structured as follows:  
\emph{Section}~\ref{sec:qr-related-work} describes the different data filtering and query rewriting strategies that are used to restrict access to RDF data.  
\emph{Section}~\ref{sec:qr-AC-correctness-criteria} presents a set of correctness criteria, which allows for access control via query rewriting to be compared to access control via data filtering.
\emph{Section}~\ref{sec:qr-beyond-simple-queries} proposes query rewriting strategies for both SPARQL 1.1 queries and updates.
In \emph{Section}~\ref{sec:qr-evaluation} we use the proposed correctness criteria to evaluate both an existing query rewriting strategy and the alternative strategy presented in this paper.
Finally, we present conclusions and directions for future work in \emph{Section}~\ref{sec:qr-conclusion}.

%%%%%%%%%%%%%%%%%%%%%%%%%
% Related Work
%%%%%%%%%%%%%%%%%%%%%%%%%
\section{Related Work}
\label{sec:qr-related-work}

%Broadly speaking access control frameworks for RDF data enforce access control either at the data layer \cite{Dietzold2006, muhleisen2010swrl}, the query layer \cite{Abel2007, Franzoni2007, Chen2009, oulmakhzoune2010, Costabello2012} or a combination of both \citet{li2008}.
%
%When access control is enforced at the data layer, a filtering mechanism is used to generate a view of the data based on a given access control policy (which only contains authorised data). 
%
%Whereas, when access control is enforced at the query layer, query rewriting techniques are used to limit access to data, based on a given access control policy. 
%
%We focus on this second approach which can be used to provide access control for the Linked Data Web by i)~specifying access control policies for one or more data sources and ii) adopting a query rewriting approach to return results using standard SPARQL.
%
%-----------------------------------------------------------
Both \citet{Dietzold2006} and \citet{muhleisen2010swrl} adopt a filtering approach to access control over RDF data.
\citet{Dietzold2006} propose access control policy specification at multiple levels of granularity (\emph{triples}, \emph{classes} and \emph{properties}). Authorisations are used to associate filters (SPARQL \texttt{CONSTRUCT} queries) with users and resources. When a requester submits a query, a virtual model is created based on the matched authorisations. The query is subsequently executed against the virtual model, which only contains data that the requester is authorised to access.
Whereas, \citet{muhleisen2010swrl} allow access control policies to be specified for \emph{triple patterns}, \emph{resources} or \emph{instances} using SWRL rules. When a requester submits a query, the system uses the access control rules to generate a temporary named graph containing only authorised data. The requesters query is subsequently executed against the temporary named graph and the results are returned to the user.  

%-----------------------------------------------------------
%In contrast, \citet{Abel2007, Franzoni2007, Chen2009} and \citet{oulmakhzoune2010} adopted a query rewriting approach to access control.
%\citet{Abel2007} and \citet{Franzoni2007} demonstrate how contextual conditions associated with the requester, the resource or the environment can be injected into the query. 
%
A number of researchers \citet{Abel2007, Franzoni2007,oulmakhzoune2012} have proposed strategies that involve adding policy information to the query (for example, the requester can only see information relating to papers that they have authored).
\citet{Abel2007} specify authorisations in terms of sets of \emph{contextual predicates}, \emph{path expressions} and \emph{boolean expressions} that are used to restrict access to instance data. 
%The authors propose a query rewriting strategy which constructs bindings for authorisation path expressions and contextual predicates. 
For positive authorisations, the path expressions and the bindings are appended to the query \texttt{WHERE} clause. For negative authorisations, the path expressions and the bindings are added to a \texttt{MINUS} clause, which in turn is appended to the query.       
\citet{Franzoni2007} also propose a query rewriting strategy which is used to grant/deny access to \emph{ontology instances}, based on contextual information pertaining to the user or the environment. 
%Authorisations are used to associate properties in the form of path expressions, attributes and filter conditions with resources.
Like \citet{Abel2007} bindings are generated for path expressions specified in the access control policy and both the path expressions and the bindings are added to the query \texttt{WHERE} clause.
%
%The approach adopted by both \citet{Abel2007} and \citet{Franzoni2007} requires that the dataset contains information that together with contextual data pertaining to the requester can be used to limit the query results.
%
\citet{oulmakhzoune2012} extend previous work by demonstrating how the \texttt{SERVICE} operator can be used to return bindings based on an external privacy preferences policy.
%
%
%The query rewriting strategies proposed by \citet{Abel2007} and \citet{Franzoni2007} are used to inject contextual information into the query as opposed rewriting the query to filter out unauthorised data. 

Both \citet{Chen2009} and \citet{oulmakhzoune2010} use simple filters to bind/unbind query solutions based on access control policies that are specified for specific \emph{classes, properties} or \emph{individuals}.
%
%\citet{Chen2009} propose a query rewriting strategy, which uses \texttt{FILTER} expressions to bind or unbind variables specified in the query to instances, properties and classes specified in the authorisations. 
According to the rewriting strategy proposed by \citet{oulmakhzoune2010}, when access is prohibited to predicates or objects, the relevant triple patterns are made \texttt{OPTIONAL}.
%, which can result in changes to the semantics of the query.     
%
%\citet{oulmakhzoune2010} also cater for both positive and negative authorisations. In the presented modelling, authorisations are composed of sets of filters that are associated with simple conditions or involved conditions. The authors use the term simple condition to refer to an authorisation, which either permits/denies access to one or more triple patterns, and the term involved condition to refer to authorisations that permit/deny access for a given \emph{predicate}.
% 
%In the case of simple conditions, when access is permitted to a single triple pattern, no action is required. Whereas, when access is denied to a single triple pattern, the triple pattern is deleted.  
%Like \citet{Chen2009} when access is permitted/denied to a basic graph pattern, the pattern is converted to an \texttt{OPTIONAL} pattern, and the authorisation \texttt{FILTER} expression is added to the query.  
%
%In the case of involved conditions, when access is permitted the \texttt{FILTER} expression, which is associated with the predicate is added to the query. If the triple pattern is not already part of the query, it is also added. When access is prohibited, the \texttt{FILTER} expression associated with the predicate is added, and if the object associated with the given predicate is a variable, a \texttt{!BOUND} expression is added for the object variable.
%
Although both \citet{Chen2009} and \citet{oulmakhzoune2010} propose query rewriting strategies for SPARQL queries, the authors focus specifically on \texttt{SELECT} queries composed of basic graph patterns and no special consideration is given to complex SPARQL queries that include subqueries or negation. 

\citet{Costabello2012} also use contextual data to restrict access to RDF. However, the proposed query rewriting strategy restricts access to \emph{named graphs} as opposed to specific classes, properties and instances. 
\citet{li2008} propose a query rewriting strategy for views generated from ontological relations. The proposed query rewriting strategy involves expanding the view concepts to include implicit concepts, retrieving both explicit and implicit access control policies and adding range and instance restrictions, associated with the matched authorisation, to the view query. 
%Like us, the authors take into account both explicit and implicit access control policies. However, they propose a combined query rewriting and filtering access control strategy, as opposed to a query rewriting strategy for the SPARQL 1.1 query and update languages in our case. 

When it comes to the evaluations of the aforementioned access control strategies only \citet{oulmakhzoune2012} examine the correctness of their access control strategy. The authors present a set of criteria which is used by \citet{Wang2007} to verify that a given access control policy holds over different relational database states and discuss how their algorithm satisfies the criteria. However, no formal evaluation is performed. %Furthermore an investigation into how the correctness criteria can be used to verify SPARQL queries that include filters is left for future work.  

In contrast to other approaches, our query rewriting strategy can be used not only to enforce access control over basic graph patterns but also in conjunction with SPARQL 1.1 queries that include subqueries and negation, graph update operations and graph management operations. In addition, we demonstrate how a set of correctness criteria, that was originally used to verify that a given access control policy holds over different database states, can be adapted to verify the correctness of access control via query rewriting, by performing a comparison with access control via data filtering. As the formal correctness criteria we propose in this paper is not specific to RDF, it can be used in conjunction with any non-monotonic query language.

\section{Access Control  Correctness Criteria}
\label{sec:qr-AC-correctness-criteria}

According to \citet{Wang2007} a query processing algorithm  should be \texttt{secure}, \texttt{sound} and \texttt{maximum}. An algorithm is \texttt{secure} if it does not return information which has not been authorised.  An algorithm is \texttt{sound} if it does not return invalid results. Furthermore, an algorithm is \texttt{maximum} if it returns as much information as possible without violating the \texttt{secure} and \texttt{sound} constraints. 
In its current form, the formal correctness criteria presented in \citet{Wang2007} is unsuitable for the verification of access control over RDF data.
Although it is possible to use their formalism to verify that a secure query processing algorithm holds over different database states, it cannot be used to verify that the algorithm is in fact secure, nor can it be used in conjunction with non monotonic queries such as \texttt{MINUS} and \texttt{FILTER NOT EXISTS}. 
%As SPARQL queries that include MINUS and FILTER NOT EXISTS may in fact be less restrictive than the corresponding query without access control, the \texttt{sound} criteria is not valid for RDF data.
%
Therefore, in this paper we redefine each of the criteria to cater for a comparison between the results obtained when:
\begin{inparaenum}[(i)]
\item a query is executed against a dataset which is generated by removing the unauthorised data  (henceforth referred to as \texttt{filtering}); and 
\item when the same query is rewritten so that unauthorised data will not be returned and subsequently executed over the unmodified dataset (which we refer to as \texttt{rewriting}).
\end{inparaenum}
%
%As RDF data does not necessarily reside in a database, the term database is taken to mean an arbitrary but fixed \emph{RDF dataset} composed of one or more \emph{RDF graphs}. 
In the definitions that follow the term \texttt{dataset} is used to denote a collection of RDF graphs, which can include a default graph and one or more named graphs. In this paper we assume that the dataset does not contain blank nodes.
Before formally defining the correctness criteria we provide definitions for an authorised dataset and an unauthorised dataset:    
\begin{definition}[Authorised Dataset and Unauthorised Dataset]
Let $D$ denote a dataset and $P$ an access control policy. Given $D$ and $P$, let $DG$ denote the set of quads in $D$ where access is granted by $P$ and $DD$ the set of quads in $D$ where access is denied by $P$. Assuming that $DG$ is disjoint from $DD$, then $DG$ is the RDF subgraph of $D$ which is authorised by $P$, and $DD$ is the RDF subgraph of $D$ which is unauthorised by $P$ .
\end{definition}

\subsection{Correctness Criteria for Non-Monotonic Queries}
\label{sec:qr-correctness-criteria-sparql-query}

%Assuming that we have two datasets, one that contains all of the data and another that only contains authorised data. 
%
%As SPARQL queries that include \texttt{MINUS} and \texttt{FILTER NOT EXISTS} may in fact be less restrictive than the corresponding query without access control, the \texttt{sound} criteria is not valid for RDF data. 
%
%As such we redefine the correctness criteria which was originally proposed by \citet{Wang2007} as follows:  
%
%A query rewriting algorithm is deemed \texttt{secure}, \texttt{sound} and \texttt{maximum} if it is not possible to differentiate between the results returned when access control is enforced using a query rewriting approach and the results returned when access control is enforced using data filtering.
%if all of the results returned by the algorithm are present in the filtered dataset. 
When access control via query rewriting is compared to access control via data filtering, a query rewriting algorithm is deemed \texttt{secure} if each of the resources returned by the algorithm are present in the authorised dataset.
The algorithm is \texttt{sound} if all of the results returned by the algorithm are also present in the result set which is generated when the original query is executed over the authorised dataset.
The algorithm is \texttt{maximum} if the data returned by the algorithm is equivalent to the data returned when the query is executed over the authorised dataset.
\noindent We formally redefine the correctness criteria proposed by \citet{Wang2007} as follows:

\begin{definition}[Query correctness criteria]
Given our definition of \emph{an authorised dataset} $DG$, let $S$ denote a query processing algorithm without access control. It follows that when query $Q$ is executed on $DG$ the result set returned by $S(DG,Q)$ only contains authorised data. 

\noindent\newline Let  $A(D,P,Q)$ represent a query processing algorithm with access control, which returns the result set authorised by $P$ when query $Q$ is executed on $D$. 
%
%\noindent\textbf{Secure.}
%For $A(D,P,Q)$ to be deemed secure, each resource $r \in  A(D, P, Q)$ must be contained in the dataset which is accessible under P.
%
A query processing algorithm is:
\noindent\newline\emph{\texttt{Secure}} if and only if
%\centerline{$\forall{_P}\forall{_Q}\forall{_D}\forall{_{PG}}\forall{_{PD}}$[$  A(D, P, Q)  \sqsubseteq  S(((D \sqcap PG) \setminus PD) , Q)$].}
%\newline\smallskip
%\centerline{$\forall{_P}\forall{_Q}\forall{_D}\forall{_{DG}}\forall{_r}$[$ r \in  A(D, P, Q)  \implies r \in DG$].}
$\forall{_P}\forall{_Q}\forall{_D}\forall{_{DG}}\forall{_r}$[$ r \in  A(D, P, Q)  \implies r \in DG$].

%\noindent\textbf{Sound.}
%
%
%To be deemed sound, \emph{A(D,P,Q)} may return less results than $S(DG,Q)$, however it should not return invalid results. 
%\noindent A query processing algorithm is 
\noindent\emph{\texttt{Sound}} if and only if
%\newline\smallskip
%\centerline{$\forall{_P}\forall{_Q}\forall{_D}\forall{_{DG}}$[$  A(D, P, Q)  \sqsubseteq  S(DG,Q)$].}
$\forall{_P}\forall{_Q}\forall{_D}\forall{_{DG}}$[$  A(D, P, Q)  \sqsubseteq  S(DG,Q)$].

%\noindent\textbf{Maximum.}
%In order to be deemed maximum the results returned by \emph{A(D, P, Q)} should be equivalent to those returned by $S(DG,Q)$. 
%\noindent A query processing algorithm is 
\noindent\emph{\texttt{Maximum}} if and only if
%\newline\smallskip
%\centerline{$\forall{_P}\forall{_Q}\forall{_D}\forall{_{DG}}$[$ A(D, P, Q)  \equiv  S(DG,Q)$].}
$\forall{_P}\forall{_Q}\forall{_D}\forall{_{DG}}$[$ A(D, P, Q)  \equiv  S(DG,Q)$].
\end{definition}

%If the desire is to maximise the information, which can be returned from the query, irrespective of the query semantics, it is possible to change the query semantics by making the unauthorised \emph{query patterns} optional, which will return additional data. 
%     
%However in doing so, using the existing evaluation criteria, our query rewriting strategy would nolonger be deemed \texttt{sound}. 
%In order to evaluate, we need to make all patterns in the query that match our authorisation \emph{OPTIONAL}, for both the query rewriting and the filtering approaches. Making it infeasible to do a direct comparison between query rewriting and filtering.      

\subsection{Correctness Criteria for Updates}
\label{sec:qr-correctness-criteria-sparql-update}

%We also define correctness criteria for both the \texttt{DELETE} and the \texttt{INSERT} operations, that can be used to verify the correctness of any SPARQL 1.1 update query. 
%
%Assuming that we have three datasets, one that contains all of the data, another that only contains authorised data, which we will refer to as the \emph{filtered dataset}, and another which contains unauthorised data. 
%
Although \citet{Wang2007} focused specifically on queries, in this paper we are also interested in updates. As SPARQL 1.1 updates change the state of the dataset, it is necessary to examine the resulting datasets as opposed to comparing the query results. 
Here we use the term \texttt{rewritten dataset} to refer to the dataset which is generated when a rewritten query is executed against a given dataset, and the term \texttt{merged filtered dataset} to refer to the merge of the unauthorised dataset with the authorised dataset after the update query has been executed.
A delete query processing algorithm is deemed \texttt{secure}, if every resource that is in the merged filtered dataset, is in the rewritten dataset.
The algorithm is \texttt{sound}, if the rewritten dataset does not contain less resources than the merged filtered dataset.
The algorithm is \texttt{maximum}, if the rewritten dataset is equivalent to the merged filtered dataset.
An insert query processing algorithm is deemed \texttt{secure}, if all of the data contained in the rewritten dataset is also in the merged filtered dataset. 
The algorithm is \texttt{sound}, if the rewritten dataset does not contain more resources than the merged filtered dataset.
The algorithm is \texttt{maximum}, if all of the rewritten dataset is equivalent to the merged filtered dataset.

\begin{definition}[Update correctness criteria]
Given our definitions for an \emph{authorised dataset} $DG$ and an \emph{unauthorised dataset} $DD$, let $U$ denote an update query processing algorithm without access control. It follows that when query $Q$ is executed on $DG$ the dataset generated by $U(DG,Q)$ only contains authorised data. 
\newpage\noindent
Let $UD$ denote a delete processing algorithm with access control, where $UD(D,P,Q)$ produces a new dataset 
%$D'$ 
when query $Q$ is executed on the subset of $D$ which is authorised by policy $P$. 
%
%\newline\newline\noindent\textbf{Secure.}
%
%For $UD(D,P,Q)$ to be deemed secure, $D'$ must contain all of the data that is present in $(U(DG,Q) \cup DD)$. 
%
%\newline\newline
\noindent A delete query processing algorithm is:
\noindent \newline \emph{\texttt{Secure}} if and only if
\newline\smallskip
\centerline{$\forall{_P}\forall{_Q}\forall{_D}\forall{_{DG}}\forall{_r}$[$ r \in  (U(DG,Q) \cup DD)  \implies r \in UD(D,P,Q)$].}
%
%\noindent\textbf{Sound.}
%
%For $UD(D,P,Q)$ to be deemed sound, $D'$ must not contain less data than $(U(DG,Q) \cup DD)$.
%
%\noindent A delete query processing algorithm is:
\noindent \emph{\texttt{Sound}} if and only if
\newline\smallskip
\centerline{$\forall{_P}\forall{_Q}\forall{_D}\forall{_{DG}}\forall{_{DD}}$[$UD(D,P,Q) \sqsupseteq  (U(DG,Q) \cup DD)$].}

%\noindent\textbf{Maximum.}
%In order for $UD(D,P,Q)$ to be deemed maximum $D'$  should be equivalent to $(U(DG,Q) \cup DD)$. 
%
%\noindent A delete query processing algorithm is 
\noindent \emph{\texttt{Maximum}} if and only if
\newline\smallskip
\centerline{$\forall{_P}\forall{_Q}\forall{_D}\forall{_{DG}}\forall{_{DD}}$[$UD(D,P,Q) \equiv  (U(DG,Q) \cup DD)$].}
\newline\newline\noindent
Let $UI$ denote an insert processing algorithm with access control and $UI(D,P,Q)$ produces a new dataset 
%$D''$ 
when query $Q$ is executed on the subset of $D$ which is authorised by policy $P$. 
%
%\newline\newline\noindent
%\noindent\textbf{Secure.}
%
%For $UI(D,P,Q)$ to be deemed secure, all of the data in $D''$  must also be present  in $(U(DG,Q) \cup DD)$.
%
\noindent An insert query processing algorithm is: 
\newline\noindent \emph{\texttt{Secure}} if and only if
\newline\smallskip
\centerline{$\forall{_P}\forall{_Q}\forall{_D}\forall{_{DG}}\forall{_{DD}}\forall{_r}$[$ r \in  UI(D,P,Q)  \implies r \in(U(DG,Q) \cup DD)$].}
%
%\noindent\textbf{Sound.}
%
%For $UI(D,P,Q)$ to be deemed sound, $D''$  must not contain more data than $(U(DG,Q) \cup DD)$.
%
%\noindent An insert query processing algorithm is 
\noindent \emph{\texttt{Sound}} if and only if
\newline\smallskip
\centerline{$\forall{_P}\forall{_Q}\forall{_D}\forall{_{DG}}\forall{_{DD}}$[$UI(D,P,Q) \sqsubseteq  (U(DG,Q) \cup DD)$].}
%
%\noindent\textbf{Maximum.}
%In order for $UI(D,P,Q)$ to be deemed maximum  $D''$  should be equivalent to $(U(DG,Q) \cup DD)$. 
%
%\noindent An insert query processing algorithm is 
\noindent\emph{\texttt{Maximum}} if and only if
\newline\smallskip
\centerline{$\forall{_P}\forall{_Q}\forall{_D}\forall{_{DG}}\forall{_{DD}}$[$UI(D,P,Q) \equiv  (U(DG,Q) \cup DD)$].}
\end{definition}

\section{Query Rewriting for SPARQL Queries and Updates}
\label{sec:qr-beyond-simple-queries}

An RDF \emph{triple} is used to asserts a binary relationship between two pieces of information. A triple is represented as a tuple $\tuple{S, P, O} \in \set{(U\cup B)} \times \set{(I)} \times \set{(I\cup B\cup L)}$, where $S$ is called the \emph{subject}, $P$ the \emph{predicate}, and $O$ the \emph{object} and \textbf{I}, \textbf{B} and \textbf{L}, are used to represent \emph{Internationalized Resource Identifier (IRIs)}, \emph{blank nodes} and \emph{literals} respectively. 
A \emph{triple pattern} is a triple that can potentially contain \emph{variables} in the subject, predicate and object positions.
A \emph{quad pattern} is a an extension of a triple pattern where the fourth element is used to denote the named graph, which can take the form of either an IRI or a variable.
In this paper, we focus specifically on rewriting the query to filter out the data which has been restricted, as such we assume that an authorisation framework has already determined the unauthorised quad patterns that are relevant for a given query. 
Although we demonstrate how access can be restricted to a quad pattern, the approach can easily be extended to work with graph patterns.
We use the \texttt{foaf} prefix to denote the FOAF Vocabulary~\url{http://xmlns.com/foaf/0.1/}~\footnote{FOAF vocabulary  Specification, http://xmlns.com/foaf/spec/.}  and the \texttt{entx} prefix to refer to our sample enterprise vocabulary~\url{http://example.org/enterprisex\#}.
The query results presented throughout this paper are based on the sample data presented in \emph{Figure}~\ref{fig:exterprise dataset} and the authorisations presented in \emph{Figure} \ref{fig:qr-rdf-policy-query}.
%Furthermore, the quad patterns presented in \emph{Figure} \ref{fig:qr-rdf-policy-query} are used to restrict the query results. 
The quad pattern, \texttt{entx:MRyan entx:salary}~\emph{?o ?g},  denies access to \texttt{May Ryan's} salary. Whereas, \texttt{entx:MRyan entx:worksFor}~\emph{?o ?g},  restricts access to information pertaining to the people that \texttt{May Ryan} works for.
%
%
%The access control policy contains an authorisation which grants access to all data and 
%
%The sample queries where executed against the data depicted in \emph{Figure} \ref{fig:exterprise dataset}, and 

\begin{figure}[!t]
\centering
\begin{lstlisting}[basicstyle=\tt\small,frame=single,numbers=left,numbersep=5pt,numberstyle=\scriptsize\color{gray}]
entx:EmployeeDetails{
entx:JBloggs rdf:type foaf:Person.
entx:JBloggs foaf:name "Joe Bloggs" .
entx:JBloggs entx:salary 60000 . 
entx:MRyan rdf:type foaf:Person .
entx:MRyan foaf:name "May Ryan" .
entx:MRyan entx:salary 33000 . 
entx:JSmyth rdf:type foaf:Person .
entx:JSmyth foaf:name "John Smyth" .
entx:JSmyth entx:salary 33000 . }  
entx:OrgStructure{
entx:MRyan entx:worksFor entx:JBloggs .  
entx:JSmyth entx:worksFor entx:MRyan .  }  
\end{lstlisting}
\caption{Enterprise Dataset}
\label{fig:exterprise dataset}
\end{figure}

\begin{figure}[!t]
\centering
\begin{lstlisting}[basicstyle=\tt\small,frame=single,numbers=left,numbersep=5pt,numberstyle=\scriptsize\color{gray},mathescape=true]
$\tuple{ \texttt{entx:MRyan},~\texttt{entx:salary},~{?o},~{?g}}$
$\tuple{ \texttt{entx:MRyan},~\texttt{entx:worksFor},~{?o},~{?g}}$
\end{lstlisting}
\caption{Unauthorised quad patterns}
\label{fig:qr-rdf-policy-query}
\end{figure}

\subsection{SPARQL Queries}
\label{sec:qr-query-rewriting-strategy}

%Prior to presenting our query rewriting algorithm, we examine the different query rewriting strategies for SPARQL graph patterns. 
The SPARQL query language supports four distinct query types (\texttt{SELECT}, \texttt{ASK}, \texttt{CONSTRUCT} and \texttt{DESCRIBE}). In each case, SPARQL graph pattern matching is used in order to determine the output of the query. Although the queries presented in this section are limited to \texttt{SELECT} queries, the proposed query rewriting strategy is identical for each of the query types.     
\paragraph{\bf{Basic Graph Patterns  and Aggregate.}} 
Basic Graph Patterns (BGPs) are sets of triple patterns and aggregates are functions that are applied to groups of solutions, for example \texttt{COUNT, SUM, MIN, MAX, AVG, GROUP\_CONCAT} and \texttt{SAMPLE}. 
When we execute \emph{Query} \ref{qu:qr-simple} without any access restrictions the identifiers, names and the salaries of all persons are returned. 

\begin{queryEnv}[Basic graph pattern]
\label{qu:qr-simple}
Given the following query:

\noindent 
\begin{lstlisting}[basicstyle=\tt\small]
SELECT ?id ?name ?salary WHERE { GRAPH entx:EmployeeDetails { 
?id foaf:name ?name. ?id entx:salary ?salary } }
\end{lstlisting}

\smallskip
\noindent The output is as follows:
\smallskip
\newline\noindent
\begin{tabular}{p{3cm} p{3cm} p{3cm}}
\hline \textbf{?id}&\textbf{?name}&\textbf{?salary}\\
\midrule
\texttt{entx:JBloggs} & \texttt{"Joe Bloggs"} & \texttt{60000}\\
\texttt{entx:MRyan} &  \texttt{"May Ryan"} &  \texttt{33000}\\
\texttt{entx:JSmyth} & \texttt{"John Smyth"} & \texttt{33000}\\
\hline
\end{tabular}
\end{queryEnv}

\noindent However, if the requester is denied access to the salary pertaining to \texttt{entx:MRyan},  by authorisation 1 in \emph{Figure}~\ref{fig:qr-rdf-policy-query}, 
we need to filter out the restricted data. This can be achieved by using a \texttt{FILTER NOT EXISTS} to filter out unauthorised data and adding it to the relevant graph pattern group.
%
%One approach would be rewrite the filter so that it eliminates the pattern specified in the authorisation from the result set. For each graph pattern group: 
%%
%\begin{enumerate}[(i)]
%\item A \texttt{FILTER NOT EXISTS} expression is generated for each quad in the graph pattern group that matches an authorisation. 
%%
%If the named graph in the query is a variable and the named graph in the authorisation is a constant, then a new graph pattern group is constructed using the named graph from the authorisation and the graph pattern from the query. Otherwise the unchanged graph pattern group is added to the filter. 
%%
%\item The constants in the subject, predicate and object positions of the authorisation are bound to the variables in the query using a \texttt{FILTER =} expression. If multiple bindings exists the \texttt{FILTER} is generated using the conjunction of the bindings. 
%%
%\item Finally, the \texttt{FILTER NOT EXISTS} is added to the relevant graph pattern group.
%\end{enumerate}
%%
\emph{Query}~\ref{qu:simple-filters} limits the result set to the identities, names and salaries of authorised persons. 

\begin{queryEnv}[Basic graph pattern restricted using binding]
\label{qu:simple-filters}
Given the following query:

\noindent 
\begin{lstlisting}[basicstyle=\tt\small]
SELECT ?id ?name ?salary WHERE { GRAPH entx:EmployeeDetails { 
?id foaf:name ?name. ?id entx:salary ?salary 
FILTER NOT EXISTS {GRAPH entx:EmployeeDetails { 
?id foaf:name ?name. ?id entx:salary ?salary 
FILTER ( ?id = entx:MRyan ) } } } } 
\end{lstlisting}

\smallskip
\noindent The output is as follows:
\smallskip
\newline\noindent
\begin{tabular}{p{3cm} p{3cm} p{3cm}}
\hline \textbf{?id}&\textbf{?name}&\textbf{?salary}\\
\midrule
\texttt{entx:JBloggs} & \texttt{"Joe Bloggs"} & \texttt{60000} \\
\texttt{entx:JSmyth} &  \texttt{"John Smyth"} & \texttt{33000} \\
\hline
\end{tabular}
\end{queryEnv}

\paragraph{\bf{Subqueries and Filters.} }

In SPARQL 1.1, negation can be achieved by filtering query results using \texttt{FILTER EXISTS}, \texttt{FILTER NOT EXISTS} or \texttt{MINUS} expressions.  Although subqueries are not classified under negation, such queries are used to limit the result set based on the results of an embedded query. 
%SPARQL subqueries are used to limit the result set based on the results of an embedded query. 
%
The following queries are constructed using an inner \texttt{SELECT} query, however the query rewriting strategy is the same for queries that contain \texttt{FILTER EXISTS}, \texttt{FILTER NOT EXISTS} and \texttt{MINUS} expressions.
\emph{Query}~\ref{qu:subqueries} returns the names of all people and the names of the people that they work for. 

\begin{queryEnv}[Subqueries]
\label{qu:subqueries}
Given the following query:

\noindent 
\begin{lstlisting}[basicstyle=\tt\small]
SELECT DISTINCT  ?employee ?manager WHERE { GRAPH ?g { 
?x foaf:name ?employee . ?y foaf:name ?manager { 
SELECT ?x ?y WHERE { GRAPH ?g { ?x entx:worksFor ?y } } } } }
\end{lstlisting}

\smallskip
\noindent The output is as follows:
\smallskip
\newline\noindent
\begin{tabular}{ p{3cm} p{3cm}}
\hline  \textbf{?employee} &  \textbf{?manager}\\
\midrule
 \texttt{"John Smyth"} & \texttt{"May Ryan"} \\
 \texttt{"May Ryan"} & \texttt{"Joe Bloggs"}\\
\hline
\end{tabular}
\end{queryEnv}

\noindent 
%If authorisation 2 in \emph{Figure}~\ref{fig:qr-rdf-policy-query} denies access to a quad in the outer query, we construct the \texttt{FILTER NOT EXISTS} expression and add it to each graph pattern group containing data which should be restricted.
As authorisation 2 in \emph{Figure}~\ref{fig:qr-rdf-policy-query} matches a quad in the inner query we add a \texttt{FILTER NOT EXISTS} to the relevant graph pattern group in the subquery. \emph{Query}~\ref{qu:subqueries-filter} results in the filtering of \texttt{May Ryan} and her manager from the result set. 
%Although it is potentially feasible to infer that \texttt{Carol} knows \texttt{Bob}, as he knows \texttt{Carol}, this information is not explicitly restricted.

\begin{queryEnv}[Subqueries restricted with binding]
\label{qu:subqueries-filter}
Given the following query:

\noindent 
\begin{lstlisting}[basicstyle=\tt\small]
SELECT DISTINCT  ?employee ?manager WHERE { GRAPH ?g { 
?x foaf:name ?employee . ?y foaf:name ?manager
{ SELECT ?x ?y WHERE { GRAPH ?g { ?x entx:worksFor ?y 
FILTER NOT EXISTS { GRAPH ?g { ?x entx:worksFor ?y
FILTER ( ?x = entx:MRyan ) } } } } } } }
\end{lstlisting}

\smallskip
\noindent The output is as follows:
\smallskip
\newline\noindent
\begin{tabular}{ p{3cm} p{3cm}}
\hline \textbf{?employee} & \textbf{?manager}\\
\midrule
 \texttt{"John Smyth"} & \texttt{"May Ryan"} \\
\hline
\end{tabular}
\end{queryEnv}

\subsection{Query Rewriting Algorithm}
\label{sec:qr-query-rewriting-algorithm}

Based on the query rewriting strategies presented in the previous section, we propose a query rewriting algorithm, which ensures that only authorised data is returned by SPARQL 1.1 queries. The algorithm takes as input a query, and a set of quad patterns that need to be filtered out of the query results, and checks each SPARQL graph pattern recursively:
\begin{enumerate}[(i)]
\item If any of the graph patterns in the outer query match any of the unauthorised quad patterns, a \texttt{FILTER NOT EXISTS} element is generated.
If the named graph in the query is a variable and the named graph in the authorisation is a constant, then a new graph pattern group is constructed using the named graph from the authorisation and the graph pattern from the query. Otherwise the unchanged graph pattern group from the query is added to the filter. 
In addition, the constants in the subject, predicate and object positions of the authorisation are bound to the variables in the query using \texttt{FILTER =} expression. If multiple bindings exists the \texttt{FILTER} is generated using the conjunction of the bindings.  
\item If any of the graph patterns in an inner \texttt{SELECT}, \texttt{EXISTS FILTER}, \texttt{NOT EXISTS FILTER} or a \texttt{MINUS} match any of the unauthorised quad patterns, a \texttt{FILTER NOT EXISTS} element is generated as described above and added to the relevant graph pattern group in the subquery or the filter expression.  
\end{enumerate}

\subsection{SPARQL Updates}
\label{sec:qr-update-rewriting-strategy}
%%
%When no access control is present, as long as the query is well formed (even if the data specified in the query does not exist) the query simply returns success. 

SPARQL 1.1 update caters for a number of update operations (\texttt{CLEAR}, \texttt{LOAD}, \texttt{INSERT DATA}, \texttt{DELETE DATA} and \texttt{DELETE/INSERT}) and a number of graph management operations (\texttt{CREATE}, \texttt{DROP}, \texttt{MOVE}, \texttt{COPY} and \texttt{ADD}). 
%
%In this section, we examine several scenarios where access to graph data is partially restricted.
In the case of update queries there are two possible options:
\begin{inparaenum}[(i)]
\item the system should inform the requester that the query cannot be completed and provide a list of the triples that cannot be deleted, inserted etc.; or
\item the system should behave as if the unauthorised data is not present.
\end{inparaenum}
If we adopt the first option, the requester will be aware that data exists which they do not have access to, and could potentially infer unauthorised information by issuing one or more additional queries. 
As a result we adopt the second option and behave as per access control via data filtering. 
For SPARQL update queries, three distinct query rewriting strategies are required. 

\paragraph{\bf{DELETE, INSERT, DELETE/INSERT.}} As the \texttt{DELETE/INSERT} operation uses graph patterns in order to determine the data to be inserted, deleted or updated, the query rewriting strategy proposed in \emph{Section}~\ref{sec:qr-query-rewriting-algorithm} is used to filter out unauthorised data.
%
%Any graph patterns in the query that match an authorisation, which denies access for that requester and the \texttt{INSERT} operation, the \texttt{DELETE} operation or both should be filtered out of the query results.
%

\paragraph{\bf{DELETE DATA and INSERT DATA.}} Given, that the \texttt{DELETE DATA} and the \texttt{INSERT DATA} operations are used to delete and insert specific data, any quads matching an unauthorised quad pattern need to be removed from the query.
%
%Any quads in the query that match an authorisation quad pattern, which denies access for that requester and the \texttt{DELETE DATA} operation, should be removed from the query. 
%
The query presented in \emph{Query}~\ref{qu:qr-SPARQL-delete-data} is used to delete all data relating to \texttt{Joe Bloggs} and \texttt{May Ryan} from the \texttt{entx:EmployeeDetails} graph. 

\begin{queryEnv}[DELETE authorised data ]
Given the following query: 

\begin{lstlisting}[mathescape, basicstyle=\tt\small]
DELETE WHERE { GRAPH entx:EmployeeDetails {
entx:JBloggs rdf:type foaf:Person.
entx:JBloggs foaf:name "Joe Bloggs" .
entx:JBloggs entx:salary 60000 . 
entx:MRyan rdf:type foaf:Person .
entx:MRyan foaf:name "May Ryan" .
entx:MRyan entx:salary 33000 . } }
\end{lstlisting}

\smallskip
\noindent If authorisation 1 in \emph{Figure}~\ref{fig:qr-rdf-policy-query} is used to prohibit the requester from deleting  \texttt{May Ryan's} salary, the query is rewritten as follows:

\smallskip
\begin{lstlisting}[mathescape, basicstyle=\tt\small]
DELETE WHERE { GRAPH entx:EmployeeDetails {
entx:JBloggs rdf:type foaf:Person.
entx:JBloggs foaf:name "Joe Bloggs" .
entx:JBloggs entx:salary 60000 . 
entx:MRyan rdf:type foaf:Person .
entx:MRyan foaf:name "May Ryan" . } }
\end{lstlisting}

\smallskip
\noindent After the rewritten query is executed over the dataset presented in \emph{Figure}~\ref{fig:exterprise dataset}, the new state of the dataset is as follows:
\newline\noindent

\begin{lstlisting}[basicstyle=\tt\small,frame=single,numbers=left,numbersep=5pt,numberstyle=\scriptsize\color{gray}]
entx:EmployeeDetails{
entx:JSmyth rdf:type foaf:Person .
entx:JSmyth foaf:name "John Smyth" .
entx:JSmyth entx:salary 33000 . 
entx:MRyan entx:salary 33000 . }  
entx:OrgStructure{
entx:MRyan entx:worksFor entx:JBloggs .  
entx:JSmyth entx:worksFor entx:MRyan .  }   
\end{lstlisting}
 \label{qu:qr-SPARQL-delete-data}
\end{queryEnv}

\paragraph{\bf{Graph based update operations.}} As the \texttt{CLEAR}, \texttt{DROP}, \texttt{ADD}, \texttt{LOAD}, \texttt{COPY} and the \texttt{MOVE} operations work at the graph level, when the requester does not have access to the entire graph these queries need to be rewritten so that they operate at the triple level.
For example the \texttt{CLEAR} operation removes all of the data from a target graph. When the requester does not have access to the entire graph, the \texttt{DELETE} operation can be used to only remove authorised data. 
\emph{Query}~\ref{qu:qr-SPARQL-clear} demonstrates how the \texttt{CLEAR} operation can be represented using a \texttt{DELETE} operation. 

\begin{queryEnv}[CLEAR authorised data ]
Given the following query: 

\begin{lstlisting}[mathescape, basicstyle=\tt\small]
CLEAR GRAPH entx:EmployeeDetails 
\end{lstlisting}

\smallskip
\noindent If authorisation 1 in \emph{Figure}~\ref{fig:qr-rdf-policy-query} is used to prohibit the requester from clearing salary information pertaining to \texttt{entx:MRyan}, the query is rewritten as follows:

\smallskip
\begin{lstlisting}[mathescape, basicstyle=\tt\small]
DELETE { GRAPH entx:EmployeeDetails { ?s ?p ?o } }
WHERE { GRAPH entx:EmployeeDetails { ?s ?p ?o 
FILTER NOT EXISTS { GRAPH entx:EmployeeDetails {?s ?p ?o
FILTER (?s = entx:MRyan && ?p = entx:salary ) } } } } 
\end{lstlisting}

\smallskip
\noindent After the rewritten query is executed over the dataset presented in \emph{Figure}~\ref{fig:exterprise dataset}, the new state of the dataset is as follows:
\newline\noindent

\begin{lstlisting}[basicstyle=\tt\small,frame=single,numbers=left,numbersep=5pt,numberstyle=\scriptsize\color{gray}]
entx:EmployeeDetails{
entx:MRyan entx:salary 33000 . }  
entx:OrgStructure{
entx:MRyan entx:worksFor entx:JBloggs .  
entx:JSmyth entx:worksFor entx:MRyan .  }     
\end{lstlisting}
 \label{qu:qr-SPARQL-clear}
\end{queryEnv}

\subsection{Update Rewriting Algorithm}
\label{sec:qr-update-rewriting-algorithm}

Based on the query rewriting strategies presented in the previous section, we propose an update query rewriting algorithm, which ensures that only authorised data is inserted and deleted.  The algorithm takes as input a query, and a set of quads that need to be filtered out of the query results. In the case of:
\begin{enumerate}[(i)]
\item \texttt{DELETE/INSERT.} The query rewriting algorithm presented in \emph{Section}~\ref{sec:qr-query-rewriting-algorithm} is used to filter out unauthorised quad patterns.
\item \texttt{DELETE DATA} and \texttt{INSERT DATA.} If any of the quads in the query match an unauthorised quad pattern these quads are removed from the query.
\item \texttt{CLEAR} and \texttt{DROP.} Negative authorisations pertaining to the specified graph are added as filters to a \texttt{DELETE} query, which is used to ensure that only authorised data is removed from the graph.
\item \texttt{ADD} and \texttt{LOAD.} Negative authorisations relating to the source and the destination graphs are added as filters to an \texttt{INSERT} query, which is used to add/load only authorised data to the destination graph. 
\item \texttt{COPY.} A \texttt{DELETE} query which is constructed using negative authorisations matching the destination graph, is used to remove all data from the destination graph. While negative authorisations matching the source and the destination graphs are added as filters to an \texttt{INSERT} query, which is used to copy only authorised data into the destination graph.
\item \texttt{MOVE.} 
%A \texttt{DELETE} query which is constructed using negative authorisations matching the destination graph, is used to remove all data from the destination graph. While negative authorisations associated with the source and the destination graphs, are added as filters to an \texttt{INSERT} query, which is used to move only authorised data into the destination graph. 
The rewriting strategy for the \texttt{MOVE} operation is the same as the \texttt{COPY} operation with an additional \texttt{DELETE} query, constructed using negative authorisations matching the source graph, which is subsequently used to remove authorised data from the source graph. 
\end{enumerate}

\section{Query Rewriting Evaluation}
\label{sec:qr-evaluation}

%According to \citet{Wang2007} an algorithm, which is used to enforce access control, should be \texttt{secure}, \texttt{sound} and \texttt{maximum}. An algorithm is \texttt{secure} if it does not return unauthorised information.  
%An algorithm is \texttt{sound} if it it does not return invalid results. Whereas, an algorithm is \texttt{maximum} if it returns as much information as possible without violating the \texttt{secure} and \texttt{sound} constraints. 
In \emph{Section}~\ref{sec:qr-AC-correctness-criteria} we formally defined a set of correctness criteria that can be used to compare access control via query rewriting against access control via data filtering. In this section, we demonstrate how the proposed correctness criteria can be used to verify the effectiveness of alternative query rewriting proposals. 
Given that SPARQL query results are dependent on pattern matching and filtering, in order to prove the correctness of a query rewriting strategy we only have to show it works for all $2^{4}$ possible combinations of quad patterns for each of the SPARQL 1.1 query types presented in \emph{Section}~\ref{sec:qr-query-rewriting-strategy} and  \emph{Section}~\ref{sec:qr-update-rewriting-strategy}.  Both the size of the dataset and the data itself are irrelevant.
%
%
%In order to evaluate our query rewriting strategy, we use this criteria to verify our query rewriting algorithms.
%, along with an unconventional approach to model checking, to perform a comparison between the two approaches.
%
%Model checking is used to verify the correctness properties of finite-state systems. The objective being, given a model of a system, exhaustively and automatically check whether this model meets a given specification. 
%
%Model checking has been successfully used to verify access control in other domains, however to date it has not been used to evaluate access control over RDF. Although we use model checking in an unconventional way.
%
%The aim of our model checking  approach is to verify the query rewriting algorithms hold, irrespective of the data, for the SPARQL 1.1 queries and updates.  
%
The entire system (test data generator, query rewriting algorithm and model checking algorithm) is implemented in Java, and the query evaluation is performed over an in memory store using Jena. The Berlin SPARQL Benchmark (BSBM) dataset generator is used to generate a dataset containing 1194 quads. The dataset, queries and authorisations used in the experiments described in this paper can be found at \url{http://correctness.sabrinakirrane.com/}.
%
%Our verification of our query rewriting strategy for the SPARQL 1.1 query language was performed as per \emph{Algorithm}~\ref{alg:modelCheckingQuery}. 
%

\subsection{Evaluation of SPARQL Query Rewriting}

In order to evaluate the different query rewriting strategies we systematically generate authorisations and queries from our auto generated dataset.The following algorithm is used to evaluate each of the auto generated queries:
\begin{enumerate}[(i)]
\item Firstly, the unauthorised quad pattern is used to remove unauthorised data, and the query is executed against the resulting authorised dataset. 
\item Secondly, the unauthorised quad pattern is used to rewrite the query based on the query rewriting algorithm and this rewritten query is executed over the dataset which contains both authorised and unauthorised data.
\item Finally, the results of both approaches are compared using the criteria presented in \emph{Section} \ref{sec:qr-correctness-criteria-sparql-query}.
% If each resource in the rewritten result set is also present in the authorised dataset, the query rewriting algorithm is \texttt{secure}. If each resource in the rewritten result set is also present in the filtered result set, the query rewriting algorithm is \texttt{sound}. Whereas, if both the rewritten result set and the filtered result set are equivalent the query is deemed \texttt{maximum}.
\end{enumerate}

\paragraph{\bf{Existing query rewriting algorithms}.}
Both \citet{Chen2009} and \citet{oulmakhzoune2010} use filters to bind/unbind query solutions based on access control policies that are associated with \emph{classes, properties} or \emph{individuals}. Although the authors propose query rewriting strategies for both positive and negative authorisations our evaluation focuses on negative authorisations.  
 
\begin{itemize}
\item \citet{Chen2009} use a \texttt{FILTER !=} expressions to remove bindings for unauthorised individuals. Whereas \texttt{OPTIONAL \{?s ?p ?o. FILTER (?p=R)\} %FILTER(! BOUND (?p))
} is used to filter matches for a named relation R and \texttt{OPTIONAL \{?s rdf:type ?o. FILTER(?o=C)\} FILTER(!BOUND (?o))} is used to filter out matches for a named class \texttt{C}.  
\item \citet{oulmakhzoune2010} also use \texttt{FILTER !=} expressions to remove bindings for specific individuals. However, according to their rewriting strategy if access is restricted to the entire triple pattern then the triple pattern is removed. Whereas, if access is partially restricted then the triple pattern is converted to an optional \texttt{OPTIONAL} pattern and the \texttt{FILTER} expression is added to the optional pattern.
\end{itemize}

\noindent As both query rewriting strategies are based on triples as opposed to quads, in order to evaluate we systematically generate authorisations from all $2^{3}$ possible combinations (of constants and variables) for each quad in the auto generated dataset (the graph is always a variable).
As the authors focus on BGPs, we auto generated queries, composed of either one, two or three RDF quad patterns that are randomly generated from the dataset, for each authorisation.    

As the query rewriting strategies proposed by \citet{Chen2009} relies on binding/unbinding constants in the authorisation to variables in the query, when the restricted class, property or individual appears as a constant in the query, it is not possible to generate a binding. In such instances their respective query rewriting strategies fail to satisfy the \texttt{secure}, \texttt{sound} and \texttt{maximum} criteria.
Whereas, according to the rewriting strategy proposed by \citet{oulmakhzoune2010}, when access is prohibited to predicates or objects, the relevant triple patterns are made \texttt{OPTIONAL}, which changes the semantics of the query. Although the query does not return any unauthorised data both the \texttt{sound} and \texttt{maximum} criteria are violated.    

\paragraph{\bf{Our query rewriting algorithm}.}
\label{sec:qr-evaluation-setting}

%The aim of our evaluation is to use this criteria to verify that the query rewriting algorithms hold for both SPARQL 1.1 queries and updates.  
%In order to evaluation our query rewriting algorithms, we developed an authorisation and query data generator, which automatically generates a set of authorisations from all $2^{4}$ possible combinations (of constants and variables) for each quad in a given dataset, and which systematically generates queries for each query type.
%%
%%, making our model checking approach unconventional. 
%
%%The benchmark system has an Intel(R) Xeon(R) CPU 8 core 2.13GHz processor, 64 GB of memory and runs Debian 6.0.3. 
%%
%The entire system (test data generator, query rewriting algorithm and model checking algorithm) is implemented in Java, and the query evaluation is performed over an in memory store using Jena. The Berlin SPARQL Benchmark (BSBM) dataset generator was used to generate a dataset containing 1194 quads. 
%
%Authorisations and queries are auto generated for the BSBM dataset, using our test data generator.
For each RDF quad in the BSBM dataset, we generate $2^{4}$ authorisations, resulting in a total of 19104 authorisations. 
As \texttt{SAMPLE} returns different data each time it is executed it is not possible to compare query rewriting to results filtering. 
Therefore, for each of the authorisations eleven queries are generated as follows: 
\begin{itemize}
\item \textbf{Basic Graph Patterns.} A query is generated, which contains either one, two or three RDF quad patterns, that are randomly generated from data selected from the entire dataset. 
\item \textbf{Aggregates.} For \texttt{COUNT} and \texttt{GROUP\_CONCAT} operations, queries are also generated from up to three RDF quad patterns that randomly generated from the entire dataset. Given \texttt{SUM}, \texttt{MIN}, \texttt{MAX} and \texttt{AVG} operations are dependent on numeric data, these queries are generated from a quad pattern which matches all offers (\emph{?s} \texttt{rdf:type} \texttt{bsbm:offer} \emph{?g}) together with a pattern that matches the associated delivery days (\emph{?s} \texttt{bsbm:deliveryDays} \emph{?o} \emph{?g}).  
\item \textbf{Subqueries and Filters.} A pattern with all variables is added to the outer query, and the inner \texttt{SELECT}, \texttt{MINUS}, \texttt{FILTER EXISTS} and \texttt{FILTER NOT EXISTS} are generated from either one, two or three RDF quad patterns, which are randomly selected from the entire dataset. 
\end{itemize}

%\begin{algorithm}[t]
% \SetAlgoLined
%\SetKwProg{myalg}{Algorithm}{}{}
%\myalg{queryModelChecking{}}
%{
% 	\KwData{dataset, auths, queries}
%	\For{Query query in queries}
%	{
%		Auth auth = getAuth(query, auths)		
%		\newline\newline
%		Dataset filteredDS = filterDataset(dataset, auth)
%			
%		List<String> filteredResults = execute(query, auth, filteredDS)		
%		\newline\newline
%		Query rewrittenQuery = rewriteQuery(query, auth)
%		
%		List<String> rewrittenResults = execute(rewrittenQuery, auth, dataset)		
%		\newline\newline			
%		secure = " secure "
%		
%		sound = " sound "
%		
%		maximum = " maximum "
%		
%		\smallskip
%		\For{String s in rewrittenResults}
%		{
%			\If{!filteredDS.contains(s)}
%			{			
%				secure= " not secure "
%			}
%
%			\If{!filteredResults.contains(s)}
%			{	
%				sound= " not sound "
%			}
%		}
%	
%		\smallskip
%		\If{rewrittenResults != filteredResults}
%		{	
%			maximum= " not maximum "
%		}
%	
%		print(query.filename + secure + sound + maximum)	
%	}
%}
%\Return	
% \caption{Query model checking algorithm}
% \label{alg:modelCheckingQuery}
%\end{algorithm}

%Based on our initial analysis, when the \emph{Quad Pattern} to be matched does not appear in the data no results are returned, however if we use a filtering mechanism to filter out the \emph{Quad Pattern} all results are returned. Further analysis is required in order to determine an appropriate rewriting strategy for these corner cases. 

\noindent In the case of basic graph patterns, aggregates and negation, each of the queries generated from the BSBM dataset were deemed \texttt{secure}, \texttt{sound} and \texttt{maximum}.
However, in the case of property paths, given a \texttt{FILTER NOT EQUALS} does not restrict access to the path data, in some instances the algorithm failed to satisfy all three criteria. If we remove the \texttt{FILTER} = binding, the query rewriting strategy is both \texttt{secure} and \texttt{sound}, however it is not \texttt{maximum}. As such, further analysis is required in order to determine a rewriting strategy for property paths.

\subsection{Evaluation of SPARQL Update Rewriting}

%\paragraph{\textbf{Verification of our update rewriting algorithm.}}

%
For SPARQL updates,the following algorithm is used to evaluate each of the auto generated queries:
\begin{enumerate}[(i)]
\item Firstly, the unauthorised quad pattern is used to create a dataset which only contains authorised data and a dataset which only contains unauthorised data. The query is subsequently executed against the authorised dataset and both the unauthorised dataset and the updated authorised dataset are merged to form a new merged filtered dataset. In the case of  \texttt{INSERT DATA} unauthorised triples need to be removed from the query before it is executed over the authorised dataset. In such instances the filtering approach is quite similar to the rewriting approach.
\item Secondly, the quad pattern is used to rewrite the query and this rewritten query is  executed over the original dataset. %Again, the the updated dataset is returned rather than a list of projections.
\item Finally, the results of both approaches are compared using the criteria presented in \emph{Section} \ref{sec:qr-correctness-criteria-sparql-update}.
%
%As \texttt{DELETE DATA}, \texttt{DELETE}, \texttt{CLEAR}, \texttt{DROP} and \texttt{MOVE} operations delete data from the dataset, if each of the quads in the filtered dataset are also present in the rewritten dataset the delete is deemed \texttt{secure}.  Whereas, if all of the data in the merged filtered dataset is present in the rewritten dataset, the delete is deemed \texttt{sound}.
%
%Given, \texttt{INSERT DATA}, \texttt{INSERT}, \texttt{ADD}, \texttt{LOAD}, \texttt{COPY} and \texttt{MOVE} operations insert data into the dataset, if all of the data in the rewritten dataset is also in the merged filtered dataset, the insert is \texttt{secure}. Whereas, if each of the quads in the rewritten dataset are also present in the merged filtered dataset, the insert is deemed \texttt{sound}. 
%
%If both the rewritten result set and the filtered result set are equivalent, both the delete and/or the insert are deemed \texttt{maximum}.
\end{enumerate}
\paragraph{\bf{Our update rewriting algorithm}.} The query rewriting strategy for \texttt{LOAD} is identical to that for the \texttt{ADD} operation and no rewriting strategy is required for \texttt{CREATE} as access is either granted or denied.    
Therefore, for each of the authorisations we generate ten different queries as follows: 
\begin{itemize}
\item \textbf{Delete Data and Insert Data.}  For \texttt{DELETE DATA} and \texttt{INSERT DATA}, queries are generated from one, two or three RDF quads, that are randomly selected from the entire dataset. 
\item \textbf{Delete, Insert and Delete/Insert.}  As per basic graph patterns, \texttt{DELETE}, \texttt{INSERT} and \texttt{DELETE/INSERT} queries are generated from graph patterns that are randomly selected from the entire dataset.
\item \textbf{Graph Update Operations.} \texttt{CLEAR}, \texttt{DROP}, \texttt{ADD}, \texttt{COPY} and \texttt{MOVE} queries are generated for each graph appearing in the dataset. 
\end{itemize}

\noindent As each of the update queries, that were generated from the BSBM dataset, were deemed \texttt{secure}, \texttt{sound} and \texttt{maximum}, we can conclude that the update query processing algorithm is \texttt{secure}, \texttt{sound} and \texttt{maximum}.

\section{Conclusion and Future Directions}
\label{sec:qr-conclusion}

Although the technology to link web data with other relevant data using machine-readable formats has been in existence for a number of years, in order to support the next generation of eBusiness applications on top of Linked Data appropriate security and privacy mechanisms need to be put in place. 
%A number of access control enforcement frameworks have been proposed for RDF, the data model which underpins the LDW. 
%However, limited research has been conducted into providing partial data access based on query rewriting. Existing query rewriting proposals do not consider complex queries that include negation or update queries. 
When it comes to access control via query rewriting, existing proposals do not consider complex queries that include negation and subqueries, and to the best of our knowledge there is currently no query rewriting strategy for update queries. 
Also, to date researchers have focused on performance evaluations, as opposed to verifying the correctness of the proposed access control mechanisms. 

In this paper, we proposed a query rewriting strategy for both the SPARQL 1.1 queries and updates. We redefined  a set of criteria, which was originally used to verify that an access control policy holds over different database states, to allow for access control via query rewriting to be compared against access control via results filtering. We subsequently used the adapted correctness criteria to evaluate the proposed query rewriting algorithms. 
In future work we plan to devise an appropriate query rewriting strategy for property paths. Based on our initial performance evaluation, which is not presented in the paper, it is evident that \texttt{FILTER NOT EXISTS} can be expensive. As such, we plan to investigate query optimisation techniques for the different SPARQL 1.1 query types and to publish a benchmark that can be used to assess alternative access control strategies.

%\subsubsection*{Acknowledgements.} 
%TBA

\label{references}

\bibliographystyle{plainnat}
\bibliography{correctness}

\end{document}